\documentclass[preprint,aps,prl]{revtex4-1}

\usepackage{graphicx}
\usepackage{amssymb}
\usepackage{amsfonts}
\usepackage{amsmath}
\usepackage{amsbsy}
\usepackage[version=3]{mhchem}  % Formula subscripts using \ce{}

\renewcommand{\vec}[1]{\mbox{\boldmath$\mathrm{#1}$}}
\newcommand{\be}{\begin{equation}}
\newcommand{\ee}{\end{equation}}
\newcommand{\ben}{\begin{eqnarray}}
\newcommand{\een}{\end{eqnarray}}

\begin{document}

\title{Chargeless spin current for switching and coupling of domain walls in magnetic nanowires}

\author{Chenglong Jia$^{1,2}$}
\email{cljia@lzu.edu.cn}

\author{Jamal Berakdar$^{2}$}
\email{jamal.berakdar@physik.uni-halle.de}

\affiliation{
$^{1}$Key Laboratory for Magnetism and Magnetic Materials of the Ministry of Education, Lanzhou University, Lanzhou 730000, China\\
$^{2}$ Institut f\"ur Physik, Martin-Luther Universit\"at Halle-Wittenberg, 06099 Halle (Saale), Germany}

\begin{abstract}
The demonstration of the generation and control of a pure  spin current (without net charge flow) by electric fields and/or temperature gradient
 has been an essential leap in the quest for  low-power consumption electronics. The key issue of whether and how such a current can be utilized to drive and control information stored in  magnetic domain walls (DWs) is still outstanding and is addressed here.
We demonstrate that pure spin current acts on  DWs in a magnetic stripe  with an effective  spin-transfer torque resulting in a mutual DWs separation dynamics and picosecond magnetization reversal. In addition, long-range ($\sim$ mm) antiferromagnetic  DWs coupling emerges.
If  one DW is pinned by  geometric constriction, the  spin current induces a dynamical  spin orbital interaction that triggers an internal electric field determined  by $\vec{E} \sim \hat{e}_{x} \cdot (\vec{n}_{1} \times \vec{n}_{2})$ where  $\vec{n}_{1/2}$ are the effective DWs orientations and $\hat{e}_{x} $ is their spatial separation vector. This leads to charge accumulation or persistent electric current in the wire.  As DWs  are routinely  realizable and tuneable, the predicted effects  bear genuine  potential for  power-saving spintronics devices.
\end{abstract}

%\date{\today}

\maketitle
%%%%%%%%%%%%%%%%%%%%%%%%%%%%%%%%%%%%%%%%%%%%%%%%%%%%%%%

\emph{Introduction}. Generally, the interaction between charge currents and localized magnetic textures, e.g. domain walls (DWs), has attracted intense research as a paradigm for the interplay of charge and spin degrees of freedom, but also  due to novel spintronic applications \cite{STT-Book,STT-Beach,Khajetoorians:2013hj}. For instance, concepts based on magnetism being controlled by currents/current-induced torques  (e.g.,  racetrack memory) are discussed as a yet new branch in the evolution of  functionalities of spintronic devices \cite{Parkin:2008gs,Thomas:2010ta,Bauer:2012fq}.  In general, current-induced magnetization dynamics can be understood  by means of the Landau-Lifshitz-Gibert equation  (LLG) \cite{Ralph:2008kj,Brataas:2012fb}. Within this phenomenological description,  two additional contributions  to the  torques may arise that stem from the  spatially nonuniform magnetization \cite{Zhang:2004hs,Tatara:2004ub}. 1) An adiabatic spin-transfer torques (STT):  in this case the carriers' spins  follow adiabatically   the  direction of the  local magnetization; and  2) a nonadiabatic contribution that gains
 importance when the carriers wave length is comparable to the spatial
extension of the non-collinear region. Thus, theoretically  the spin-polarized,
 charged current-induced DW dynamics is fairly  well understood. For applications, however, obstacles have to be circumvented that
 are associated by the required
  high charge current density entailing  high  energy consumption and dissipation.
%
%
%
%%%%%%%%%%%%%%%%%%%%%%%%%%%%%%%%%%%%%%%%%%%%%%%%%%%%%%%%%
\begin{center}
\begin{figure}[b]
%\centering
%\subfigure[]
{
\begin{minipage}[]{0.9\textwidth}
\begin{center}
\includegraphics[width=0.8\textwidth]{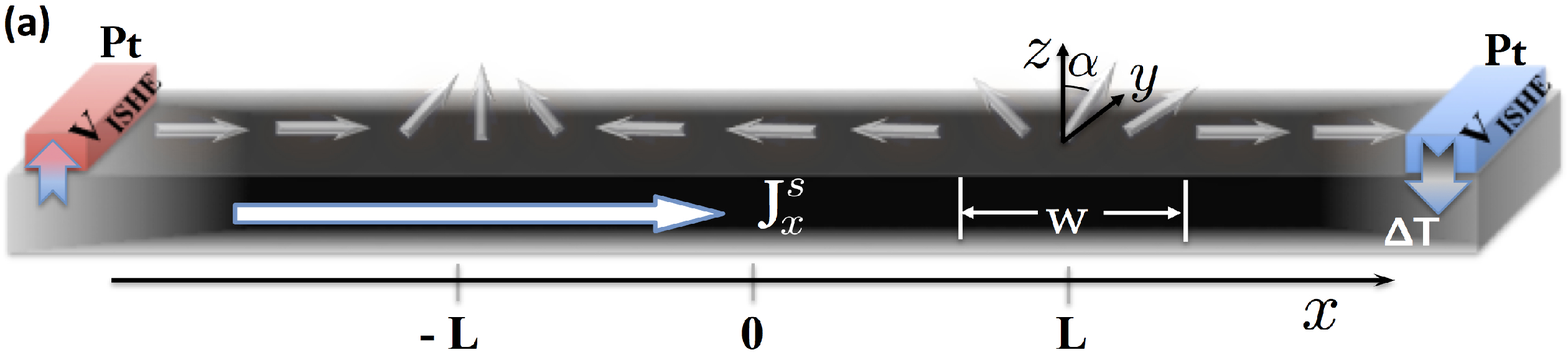}
\end{center}
%\\
%
\includegraphics[width=0.45\textwidth]{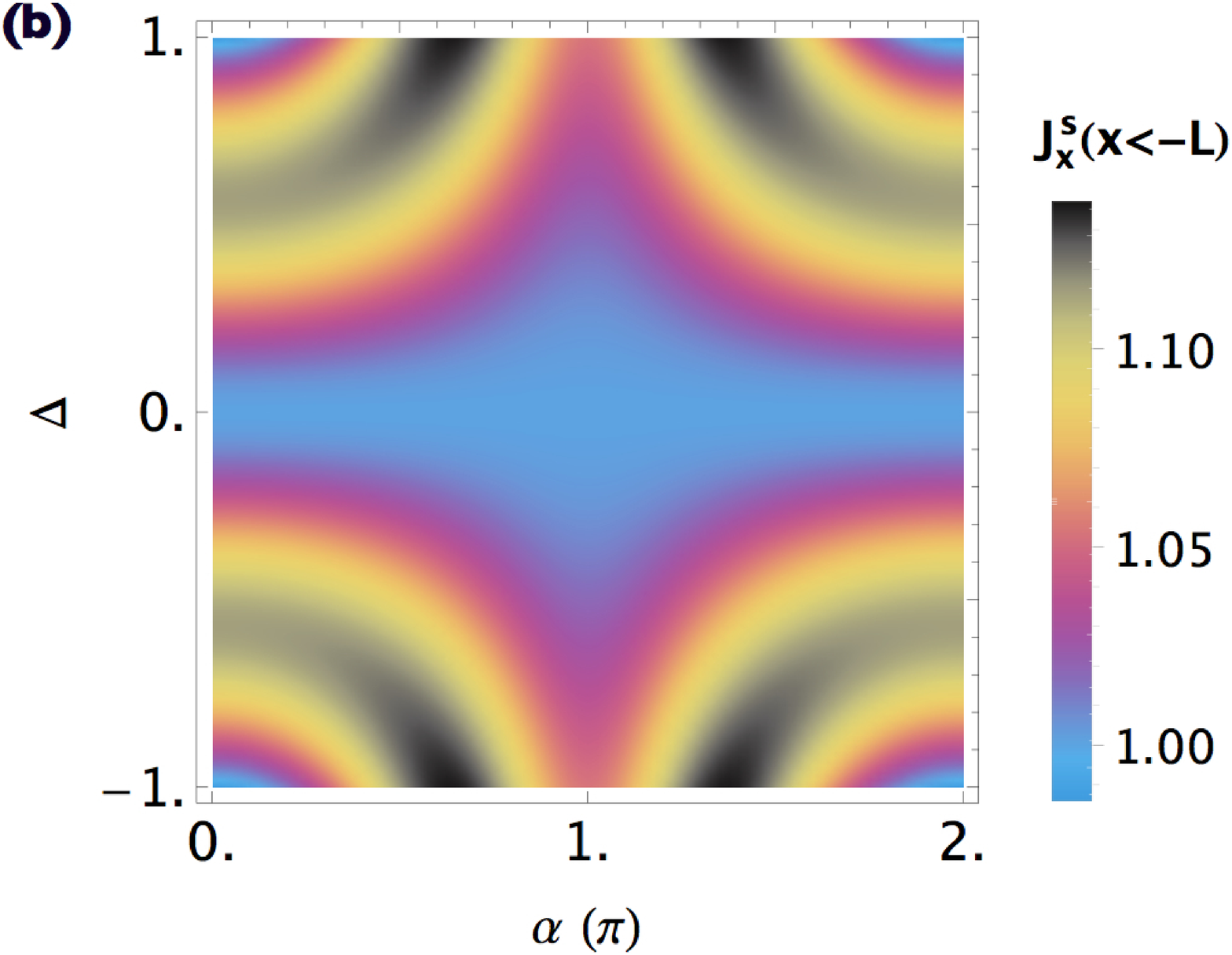}
\includegraphics[width=0.45\textwidth]{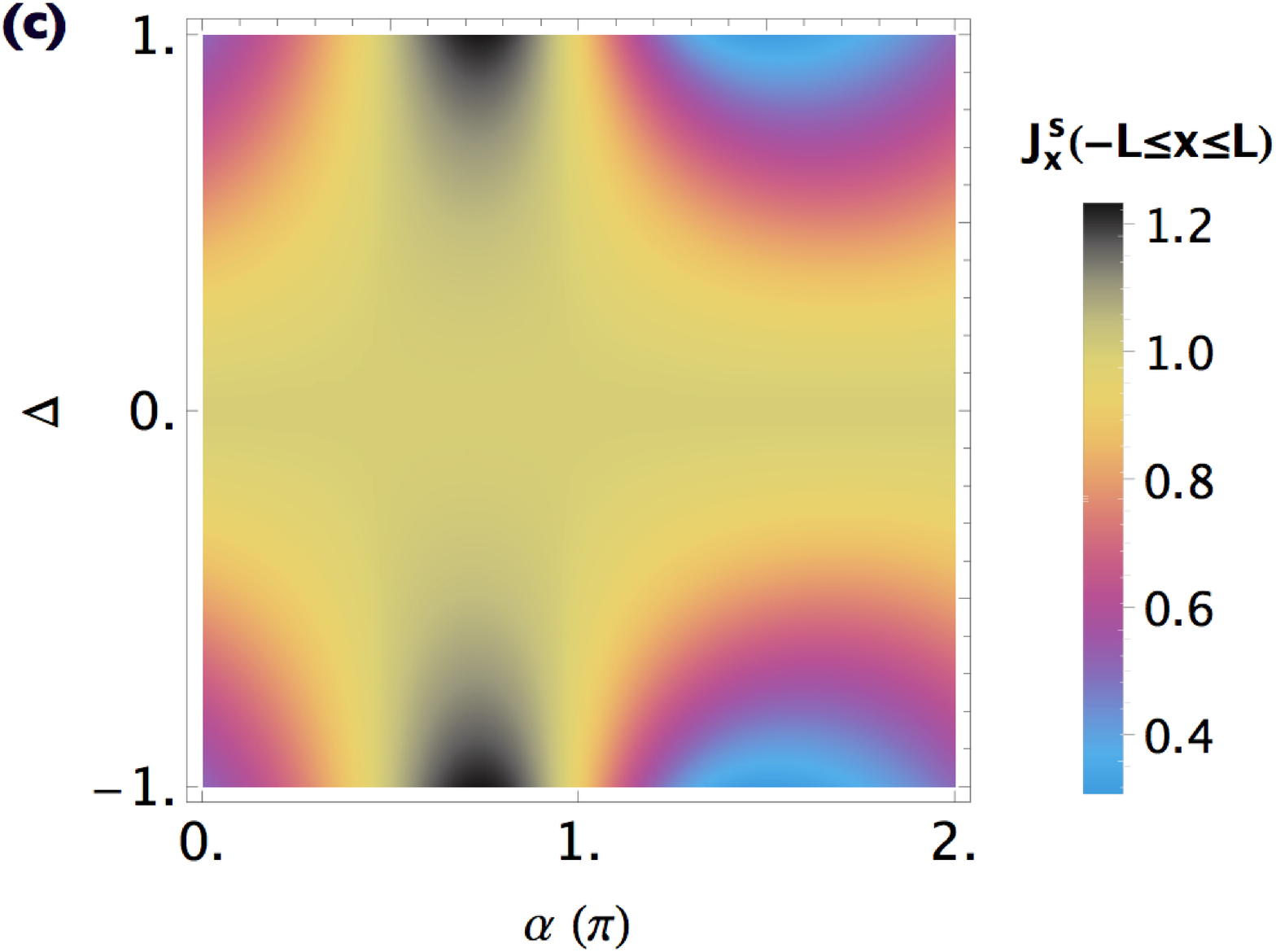} \\
\includegraphics[width=0.45\textwidth]{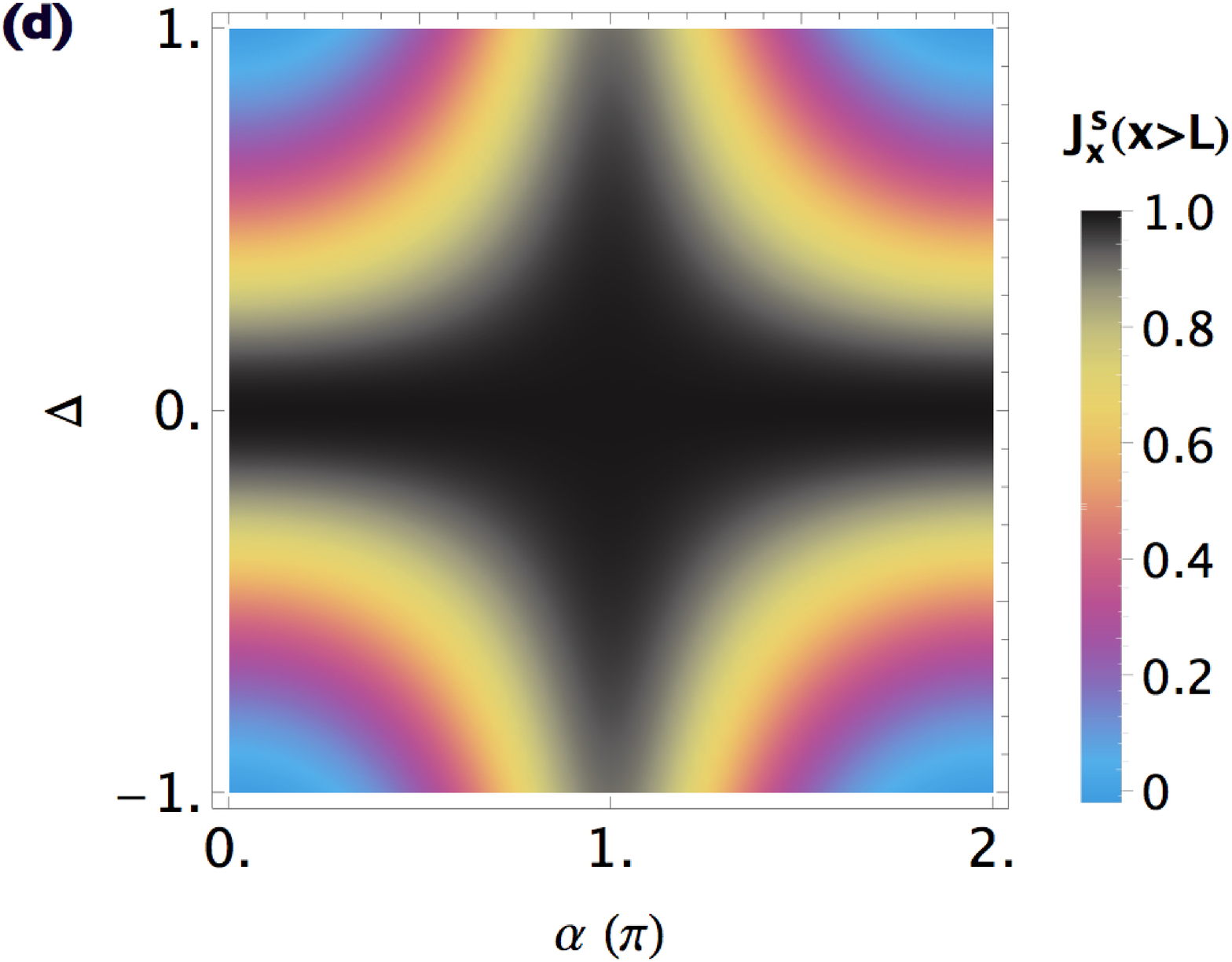}
\includegraphics[width=0.45\textwidth]{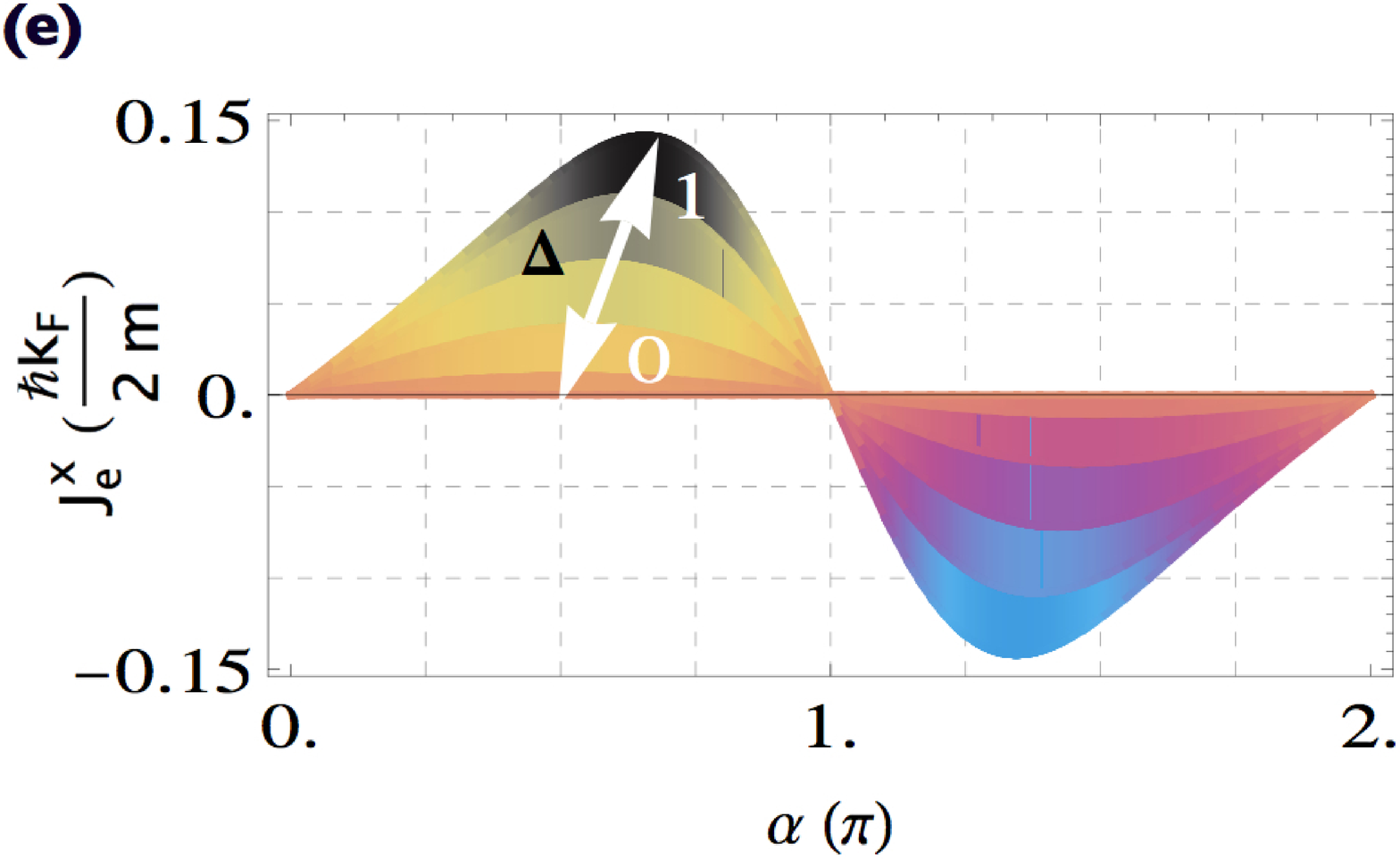}
\end{minipage}
}
\caption{(a) A ferromagnetic wire containing double domain walls subject to a spin current $\vec J_{x}^{s}$, which can be induced by  the spin injection and/or spin pumping due to a mismatch of  spin electrochemical potential with the Pt strips deposited on the sample, e.g., through the inverse spin-Hall effect (ISHE) or the longitudinal SSE. The magnetization profile is as shown schematically (thick arrows). $\pm L$ and $\alpha$ are, respectively, the DWs positions and orientation with respect to the  $xz$-plane. $w$ is the DW width. (b-c) Inhomogeneous spin current in the wire. (e) The magnetic scattering induced charge current with $C_{2}$-symmetry. $\Delta$ is the effective spin-current scattering strength.
 }
\label{Fig::system}
\end{figure}
\end{center}
%%%%%%%%%%%%%%%%%%%%%%%%%%%%%%%%%%%%%%%%%%%%%%%%%%%%%%%%%
%
%
%
%

As shown below, a surprisingly effective solution to these problems is achieved by going in a qualitatively novel direction and utilizing
  a pure carrier spin current,  i.e.  a flow of electron spin angular momentum.
  That such a spin current can be generated in a versatile and energy-effective way is
   evidenced by an impressive series of recent discoveries:
 E.g., an open circuit spin currents can be  generated by a temperature gradient in a spin-Seebeck effect  (SSE) geometry \cite{SSE-metal,SSE-inst,SSE-semi,SSE-nom}, or due to spin-Hall effect \cite{Kajiwara:2010uc}, or  by means of ferromagnetic resonance \cite{Shikoh:2013jp}.  These findings open the way for spintronic devices controlled by  spin flow with the advantage of energy-consumption reduction \cite{Book-SpinCurrent}.  The issue of how a long-lived \emph{pure} carrier spin current can be utilized to steer  magnetic textures (DWs)
 is obviously of a critical importance and has not yet been addressed, to the best of our knowledge.
 %For a magnetic moment embedded in a conductive magnetic wire, it is shown that the SSE-induced spin current may scatter from this localized nanostructure, setting it in a %precessional and displacement motion \cite{SSE-ansatz}.
%On the other hand, as the density of localized magnetic moments increase and/or DWs are formed, it is necessary to consider the effects of the current-induced interactions on the magnetic texture. Reference [\citenum{DWs-Nich}] pointed out an occurrence of a RKKY-like electron-current-mediated interaction of multiple DWs.
In this Letter, we show that in a magnetic nanowire with double DWs (cf. Fig.\ref{Fig::system}a),  interferences due to
spin-dependent transmissions and  reflections of the spin current  give rise to an indirect long-range antiferromagnetic coupling between DWs.  An \emph{intrinsic} electric field related to the DWs configuration with a broken mirror symmetry is predicted, \be \vec{E} \sim \hat{e}_{x} \cdot (\vec{n}_{1} \times \vec{n}_{2}) \ee where $\vec n_{1}$ and $\vec n_{2}$ denote the directions of DWs polarization, respectively. By pinning one of the DWs at a constriction, the second DW is found to possess  an ultrafast ($\sim ps$) magnetic reversal and inter-domain wall displacement, evidencing that indeed spin current is a qualitatively new tool for spin-dynamics control. \\

\emph{Model}.  The system under consideration is illustrated in Fig.\ref{Fig::system}a: a ferromagnetic (FM) wire with a magnetization profile exhibiting two DWs, down which a spin current, $\vec J_{x}^{s}$, is passed.  The two DWs have the same extension $w$ and are separated by the distance  $2L$. Their profile $\vec{M}(x) = \text{M}_{0} \vec{n}(x)$ is parameterized by the angles $\alpha(x)$ and $\varphi (x)$, i.e., $n_x = \cos \varphi $, $n_y = \sin \alpha \sin \varphi$, and $n_{z} = \cos \alpha \sin \varphi$, where
%
%\be
$\varphi(x) = \arccos \left[ \tanh \frac{x+L}{w} \right] +\arccos \left[ \tanh \frac{x-L}{w} \right].$
%\ee
%
Without loss of generality, we set $\alpha(-L) = \alpha_{1} = 0$  at the first wall and  $\alpha(L) = \alpha_{2} = \alpha$ around the second.  The interaction between the delocalized electronic states with spin $\sigma$
 and the (classical) localized moments forming the DWs are modeled by the ``\emph{s-d}'' Hamiltonian \cite{Zhang:2004hs,DWs-JB06} ($g$ is a local exchange coupling strength)
\be
   H_{sd} = g \vec{M}(x) \cdot \vec{\sigma}.
\label{eq:hint} \ee
Independent of the different underlying mechanisms for generating the spin current density, $\vec{j}_{\mu}^{s}(k)$, we can describe it with  chargeless eigenstates $\psi_{B} (x)$ \cite{SSE-ansatz},
\be
   \psi_{B} (x) =\frac{1}{2} \left[ e^{ikx} \binom{1}{1} + e^{-ikx} \binom{e^{i\theta}}{e^{-i\theta}}  \right].
\label{eq:psi}
\ee
The first term generates a spin current along the wire; $\theta \in [0, 2\pi]$ in the second term accounts for residual spin precession and diffusion. With this, one obtains a finite spin expectation value
 along the wire only, i.e. $\langle \sigma_x \rangle \neq 0$ but $\langle \sigma_y \rangle = 0$.  The $x$ expectation value of the charge current $ \vec{j}^e (k) = \frac{i\hbar}{2m} [ (\vec{\nabla} \psi^{\dag} (x)) \psi(x) - \psi^{\dag}(x) \vec{\nabla} \psi (x) ]$ vanishes, i.e. $j^e_x (k) \equiv 0$ (here $m$ is the effective mass).
In contrast, for the spin current $ \vec{j}_{\mu}^s (k) = \frac{i\hbar}{2m} [ (\vec{\nabla} \psi^{\dag} (x))  \sigma_{\mu} \psi(x) - \psi^{\dag}(x) \sigma_{\mu} \vec{\nabla} \psi (x)]$ one finds $J^s_{y} = \oint j^s_y d\theta/2\pi = 0 $, whereas $J^s_x = J^s_0 = \hbar k/2m$, respectively. These properties are particularly in line with experimental observations in the SSE geometry \cite{SSE-metal,SSE-inst,SSE-semi,SSE-nom}.

The primary quantity on which we focus our interest will be the spin-current transmission/reflection mediated DWs coupling. If the internal structure of $\vec M(x)$ varies on a scale larger than the electron wavelength at the Fermi surface $k_{F}$, then the spin of conduction electrons  follows  the smoothly varying nonlinear magnetic texture. One can then unitarily transform to align locally with $\vec M$, which introduces a nontrivial Berry curvature field and gives rise to the ``spin motive force'' \cite{Barnes:2007ko,Yang:2009ko,Schulz:2012bi}. On the other hand,  for sharp DWs, i.e. if $\lambda_F = 2\pi/k_F \gtrsim w$ (as realized in magnetic semiconductor-based structures),
 DW scatters  strongly  the carriers acting in effect as $\vec{M} (x) = \vec{M}_{0} \delta(x)$ \cite{DWs-JB06,Dugaev:2006ed}. Consequently,  the scattering spinor wave function in the presence of double DWs can be expressed as,
\be
\psi'_{s}(x) =
  \begin{cases}
     \frac{e^{ikx}}{2} \binom{1}{1} + \frac{e^{-ikx}}{2} \binom{r_1}{r_{1}'} + \frac{e^{-ikx}}{2}\binom{t_{1} e^{i\theta}}{t_{1}' e^{-i\theta}} ~~~ x < -L,  \\
     \frac{e^{ikx}}{2} \binom{t_{2}}{t_{2}'} + \frac{e^{-ikx}}{2} \binom{r_{2}}{r_{2}'} + \frac{e^{-ikx}}{2}\binom{t_{2} e^{i\theta}}{t_{2}' e^{-i\theta}}+ \frac{e^{ikx}}{2}\binom{r_{2} e^{i\theta}}{r_{2}' e^{-i\theta}},    \\  ~~~ ~~~ ~~~~~~ ~~~ ~~~ ~~~ ~~~ ~~~~~~ ~~~ ~~~~~~ ~~~ ~~~  ~~~ |x| \le L,  \\
      \frac{e^{ikx}}{2} \binom{t_1}{t_{1}'} + \frac{e^{-ikx}}{2}\binom{e^{i\theta}}{e^{-i\theta}} +\frac{e^{ikx}}{2}\binom{r_{1} e^{i\theta}}{r_{1}' e^{-i\theta}}    ~~ x > L.
    \end{cases}
  \label{spinor}
\ee
The complex coefficients $t_{i}$($t'_{i}$) and $r_{i}$($r'_{i}$) describe the spin reflection and transmission with reference to the original spin and spin-flip channels, and they can be analytically determined by the wave function continuity at $x = \pm L$.

\emph{Magnetoelectric effect}. At first, let us inspect the case of a small Fermi wave vector $k_{F}$;  or $k_{F} L \ll 1$. We have then $t_{1}' = t_{1}^{\star} = \frac{i}{i + \Delta}$ and $r_{1}' = r_{1}^{\star} = -\frac{\Delta}{i + \Delta}$ with $\Delta = kw \times \frac{gM_0}{\hbar^2 k^2/2m} $ being an effective spin-current scattering strength.  We readily infer  that the spin current is not influenced by the detailed configuration of two DWs and scatters equivalently from a composite magnetic cluster, setting such a collection of localized moments in a precessional and displacement motion as discussed in Ref.[\citenum{SSE-ansatz}].   For a intermediate Fermi wave vector and/or a sufficiently large distance between the DWs, such that $k_{F} L \geq 1$ whereas $k_{F} w \ll 1$,  the spin-dependent interferences due to scattering from DWs result in   a long-range interaction of  DWs.  Here the spin coherence is necessary which seems to be realizable ($L_s \sim mm$) via thermal \cite{Kajiwara:2010uc,Uchida:2012ew} or dynamical \cite{Shikoh:2013jp} approaches. Physically, the scattering leads to a redistribution of the spin electrochemical potential along the wire, resulting in a non-conserved spin current. Its density is inhomogeneous in three different region (cf. Fig.\ref{Fig::system}(b-d)), which can be imaged
%can be reflected by the polarity-reversal symmetry break of a charge voltage that emerges,
electrically by means of  the inverse spin Hall effect, e.g. by measuring a voltage build up in a
 Pt strip deposited on the sample perpendicular to the transport of spin current. Importantly, quite different to  a short pseudocircuit in the charge channels in the case of single magnetic scattering \cite{SSE-ansatz}, now we have a spin-current-induced persistent electric current
\begin{widetext}
\be
\langle j_{x}^{e} \rangle = - \frac{\hbar k}{2m} ~ \frac{32 \Delta ^2 \sin \alpha  \sin (4kL)}{\left[\Delta ^4-4 \Delta ^2 \cos \alpha - \left(\Delta ^4+4 \Delta ^2+8\right) \cos (4kL) \right]^2+16 \left(\Delta ^2+2\right)^2 \sin ^{2} (4kL)}
\ee
\end{widetext}
after integrating out the residual spin precession and diffusion over $\theta$ (see also in Fig.\ref{Fig::system}e). Phenomenologically, the DW has been predicted to behave at low energy as a magnetic impurity \cite{Saitoh:2004fa},  and consequently we have now two localized  magnetic moments indirectly coupled through the carrier-mediated exchange interaction. Given $\alpha \neq 0 (or ~\pi)$, from the symmetry point of view, the DWs configurations loose the mirror symmetry with respect to the $zx$ plane, which points to an antisymmetric Dzyaloshinsky-Moriya interaction along the wire \cite{Book-Yosida}.
In fact, the $E_{x}$ component of an electric field is allowed by symmetry, as discussed for multiferroic transition metal oxides \cite{JOHN}. Qualitatively, such an internal electric field is expressible by the vector products of DWs, $E_{x} \sim \hat{e}_{x} \cdot (\vec{n}_{1} \times \vec{n}_{2})$ that would induce charge accumulation and/or electric current with $C_{2}$ symmetry, as demonstrated in Fig.\ref{Fig::system}e. Note that the angular-dependence of the induced charge current  rules out a possible $C_{2v}$ planar Nernst effect \cite{p-NE-1,p-NE-2}. Thus, we can identify the rearrangements of DWs upon  the encounter with the spin current having
  a defined direction as the underlying mechanism for  the emergence of the electric field along the wire. This is insofar important as,  beside the spin-current-induced macroscopic magnetoelectric effect, the setup in Fig.\ref{Fig::system}a possesses multiple functionalities: it can be a thermally driven electric generator \cite{Bauer:2010iv} provided that two DWs  are mechanically pinned in non-coplanar manner.
%%%%%%%%%%%%%%%%%%%%%%%%%%%%%%%%%%%%%%%%%%%%%%%%%%%%%%%%%
\begin{figure}[b]
\centering
\includegraphics [width =0.85 \textwidth]{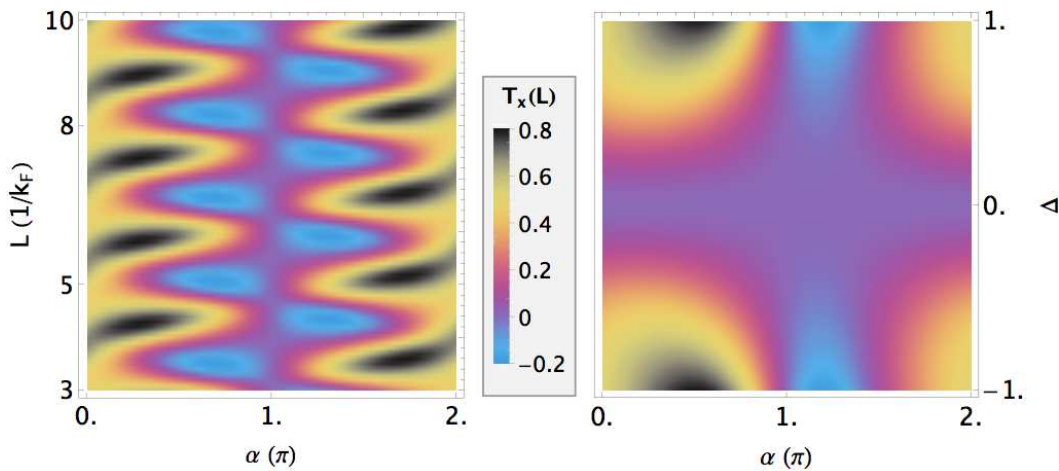}
\caption{The $\hat{x}$ component of the spin-current-induced  STT  $\vec{T}_{x}(L)$ around the right domain wall as functions of  the polar angle $\alpha$ and  (\emph{left}) the DWs distance $L$ with $\Delta = 0.72 $, or (\emph{right}) the effective spin-current scattering strength $\Delta$ with $k_{F}L = 3$, respectively.  }
\label{Fig::JTs}
\end{figure}
%%%%%%%%%%%%%%%%%%%%%%%%%%%%%%%%%%%%%%%%%%%%%%%%%%%%%%%%%

%%%%%%%%%%%%%%%%%%%%%%%%%%%%%%%%%%%%%%%%%%%%%%%%%%%%%%%%%
\begin{figure}[b]
\centering
\includegraphics [width=0.95 \textwidth]{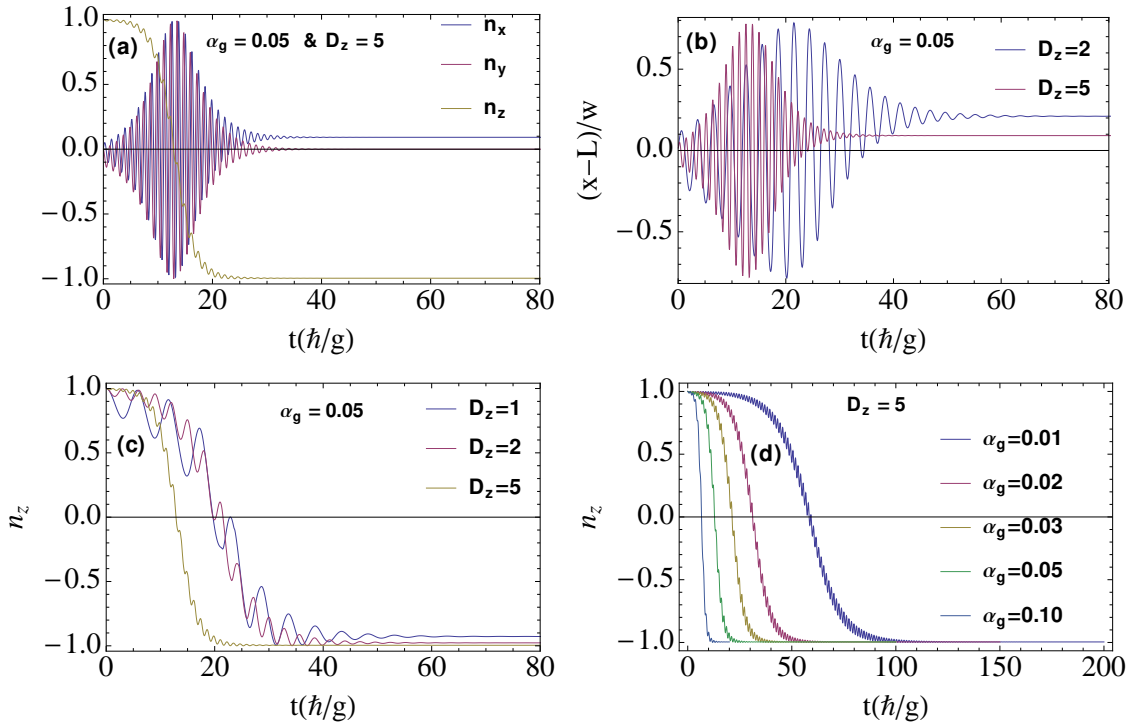}
\caption{The magnetization dynamics of the right domain wall: (a) Magnetic reversal. (b) Oscillation and displacement of the DW. The spin-current-induced magnetic reversal with respect to the transition from parallel to anti-parallel DWs configuration for different (c) perpendicular anisotropy $D_{z}$ (in scale of $g$) and (d) Gilbert damping $\alpha_{g}$.  For the numerical calculations we take the spin-current scattering strength $\Delta = 0.72$. }
\label{Fig::DW-t}
\end{figure}
%%%%%%%%%%%%%%%%%%%%%%%%%%%%%%%%%%%%%%%%%%%%%%%%%%%%%%%%%

\emph{Magnetic dynamics}. The interference of incoming and reflected waves are expected to result in  current-induced toques acting on the walls \cite{Brataas:2012fb,Ralph:2008kj}. Subsequently, we study  STT and DWs dynamics. For definiteness, let's assume that one of the DWs (say left) at $x=-L$ is pinned, and concentrate on the effect of spin current on the right DW with initial magnetization $\vec{M}$ perpendicularly polarized at $x=L$ (i.e., $\alpha(t=0) =0$). The spin current acts on the right DW with a torque $\vec{T}_{\mu}(x)$ that follows  from the jump in the spin current at the point $x$, or equivalently, within our model, from the nonequibrium spin density $s_{\mu}(x)$ accumulation
\be
\vec{T}_{\mu}(x) = - \frac{gM_{0}}{\hbar}[ \vec n \times \vec{s}(x)]_{\mu},
\label{eq:STT}
\ee
where $\vec{n}$ is the unit vector along $\vec{M}$, and  the  spin density we obtain from $s_{\mu} (x) = \psi_{s}^{\prime \dag}(x) \sigma_{\mu} \psi'_{s}(x)$. Upon scattering, the spin current carried by Eq.(\ref{eq:psi}) is modified, nonzero $J_{y}^{s}$ and $J_{z}^{s}$ emerge inhomogeneously  in three different regions: $x < -L$, $|x| \leq L$, and $x > L$. The calculated STT on the right DW is shown in Fig.\ref{Fig::JTs}. Clearly, the STT depends on the sign of the current and periodically on the DWs relative angle $\alpha$ and distance $2L$. The magnetization dynamics are then modelled with the modified Landau-Lifshitz-Geilbert (LLG) equation \cite{Brataas:2012fb,Bazaliy:1998vo}
%
%
%Under a spin current, a uniform ferromagnetic state is unstable. Rather,  a possible ground state is a state containing domain walls %\cite{D-nucleation}.
%
\begin{equation}
  \frac{\partial \vec{n}}{\partial t} = \frac{D_z}{\hbar} [\vec{n} \times \hat{e}_z] + a_g \vec{n} \times \frac{\partial \vec{n}}{\partial t} -\frac{g}{\hbar} [\vec{n} \times \vec{s}(L)]
  \label{LLG}
\end{equation}
where  $a_g$ is the Gilbert damping parameter and $D_z$ is the perpendicular anisotropy energy \cite{Miron:2011fn,Curiale:2012iq}.  It should be noted that the temperature effect can be included as a stochastic field contributing to the effective magnetic field in LLG equation, however, it is shown that at low temperature, thermal fluctuations do not alter qualitatively the LLG dynamics \cite{Alexander:2009rl}.  With an initial parallel configuration of the DWs, we calculated the time dependence of the right DW magnetization shown in Fig.\ref{Fig::DW-t} \cite{footnt}.  As concluded from the figures, the right DW is set in oscillating motion immediately when subjected to the spin current, which is different from the continuous displacements of a single localized magnetic structure \cite{SSE-ansatz,Li:2004kh}.  Furthermore, in spite of the large deviation of the spin density from the localized magnetization, the non-adiabatic torque \cite{Zhang:2004hs} is absent in our case and the DW motion is terminated with  magnetic reversal and small center shift of the wall along the wire. During its propagation,  STT causes  DW distortion  developing an out-of-plane component of the magnetization, which exhibits a fast oscillation mode exacerbated by the anisotropically damped motion in the $\hat{z}$ direction (c.f. Fig.\ref{Fig::DW-t}(b-c)). The magnetization switching time  can be shown to be mainly determined by $D_{z}/\alpha_{g}$. Even a small spin-current scattering strength, $\Delta =0.01$ is found to give a magnetic reversal because of a substantial reduction of the critical current due to the perpendicular anisotropy \cite{Jung:2008kq}.

\emph{Discussion}.  On the scale of the carriers motion  DW may be considered nearly static. Let us assume that two DWs are located at their potential minima as local spins, $\vec{M}_{1}$ and $\vec{M}_{2}$. The spin-current channel naturally gives rise to an effective multi-orbital  electron hopping between the spins $\vec{M}_{i}$. The Hamiltonian describing it, can take a simple form $H_{t} = - t \sum_{ia,jb} c_{jb}^{\dag}c_{ia}$ with $t$ being the amplitude of hopping from the orbital $a$ on the site $i$ to the orbital $b$ on the site $j$. Given that the time of the magnetic reversal ($ t \sim 10 \hbar g^{-1}$) is much longer than the hopping time ($\sim \hbar g^{-1}$), to a lowest order we obtain an effective antiferromagnetic interaction between  DWs \cite{Mostovoy:2011dz}, \be H_{S} = \frac{t^{2}}{g/4} ( \vec{n}_{1} \cdot \vec{n}_{2} +1). \ee Considering the macroscopic spin coherence length over $mm$, one has thus a long-range spin-current induced antiferromagnetic coupling, resulting in anti-parallel configuration of two DWs in equilibrium state, which is consistent with the long-time behavior of magnetic dynamics  (cf. Fig.\ref{Fig::DW-t}). Furthermore, there should be a dynamic magnetoelectric coupling during the distorted DWs oscillations as well \cite{Barnes:2007ko,Yang:2009ko}, which allows  possibly for a dynamical electric control of DWs structures by  voltage pulse.

{\emph{Conclusions and outlook}}. Pure spin current interaction with magnetic DWs  results in 1.) long-range antiferromagnetic DWs coupling,
 % If  the exchange coupling between the carriers and localized DWs to be on the scale of $meV$,
 %
and 2.) in STT-induced picosecond magnetic reversal for  DWs with  initially parallel  polarizations. 3.)
For two DWs  pinned in a non-coplanar configuration, an internal magnetoelectric effect with $C_{2}$ symmetry is predicted. \\
Linking theory to experiment we note:
favorable materials satisfy  $w < \lambda_F<L$ predestinating dilute magnetic semiconductors \cite{metals}.
E.g., for GaMnAs the domain wall width, $w$ varies within $4-12$ nm \cite{Gourdon:2007ko} and the hole concentration is around $10^{18} - 10^{20}$ cm$^{-3}$ \cite{Matsukura:1998tx}. We estimate  $k_{F} = \pi \rho_{1D} = S\pi \times 10^{-2}$ nm$^{-1}$, where $\rho_{1D}$ is the linear hole density related to the bulk density by $\rho_{1D}= \rho_{3D} S$ and $S$ is the cross section (in  unit of nm$^{2}$) and $\rho_{3D} = 10^{19}$ cm$^{-3}$. Thus, the area of the cross section is      roughly $\sim  10$ nm$^{2}$.
The local exchange coupling  $g$ in GaMnAs deduced from  experiments \cite{Goennenwein:2003kl,Haghgoo:2010ik}  is about $0.02 - 1$ meV;  the perpendicular anisotropy energy density varies (also with temperature) in $0.01-0.05$ meV/nm$^{3}$ \cite{Gourdon:2007ko,Dourlat:2008gw,Haghgoo:2010ik}. The damping coefficient $\alpha_{G}$ as deduced from the domain wall mobility measurement \cite{Dourlat:2008gw,Adam:2009iu} is $\alpha_{G} = 0.25 \pm 0.05$ (from ferromagnetic resonance  $\alpha_{G} \approx 0.01$ \cite{Khazen:2008fw,Cubukcu:2010cy}). In GaMnAs  pure spin-current  may be delivered by  SSE  \cite{SSE-semi}.
Thus, all parameters in the present theory and simulations are experimentally accessible.
The theory is also applicable to open circuit  dynamics in magnetic textures in insulators or molecules coupled to
 a spin current, e.g.
LaY$_{2}$Fe$_{5}$O$_{12}$/Fe  \cite{SSE-inst}, or FM/Mn$_{12}$/Fe \cite{Mn12}.
 %
%
%
%      pure spin-current-induced effects electrical/dynamical approaches for generating long-lived spin currents are favorable.% , since thermal fthermal  methods \cite{Bauer:2010iv,Yuan:2010hb,Hals:2010eu} is important and the.
%The longitudinal SSE configuration (cf. Fig.\ref{Fig::system}a)  can also be used to produce pure spin current in a semiconductor sample.  Another alternative to have long-range spin current via thermal method is to take a non-magnetic (InSb as in Ref.\citenum{SSE-nom}) instead of ferromagnetic material as the sample, in which  case  large spin currents can be induced by a small temperature gradient due to the giant SSE and   the effects of heat currents may be ignored.
 Hence, the present  approach is versatile  and allows to
  uncover spin-current-induced  effects such as magnetoelectric effect and ultrafast DW reversal  pointing so to
  new possible concepts for efficient spintronic devices.

\acknowledgements{This work was supported by the National Basic Research Program of China (No. 2012CB933101), the German Research Foundation (No. SFB 762, and BE 2161/5-1), the National Natural Science Foundation of China (No. 11104123), the Program for Changjiang Scholars and Innovative Research Team in University (No. IRT1251), and the Fundamental Research Funds for the Central Universities (No. 2022013zrct01/2022013zr0017). J.B. acknowledges
useful discussions with S.S.P. Parkin on the experimental aspects. }

%%%%%%%%%%%%%%%%%%%%%%%%%%%%%%%%%%%%%%%%%%%%%%%%%%%%%%%%%

%%%%%%%%%%%%%%%%%%%%%%%%%%%%%%%%%%%%%%%%%%%%%%%%%%%%%%%%%

%%%%%%%%%%%%%%%%%%%%%%%%%%%%%%%%%%%%%%%%%%%%%%%%%%%%%%%%%%%%%%%%%%%%%%%%%%%%%%%%%%%%%%%


\begin{thebibliography}{99}

\bibitem{STT-Book} A. Thiaville and Y. Nakatani, \emph{Spin Dynamics in Confined Magnetic Structures} III, Topics in Applied Physics Vol. \textbf{101} (Springer-Verlag, Berlin, 2006).

\bibitem{STT-Beach} G. S. D. Beach, M. Tsoi, and J. L. Erskine, J. Magnetism and Magnetic Materials \textbf{320}, 1272 (2008).

\bibitem{Khajetoorians:2013hj} A. A. Khajetoorians, B. Baxevanis, C. Hubner, T. Schlenk, S. Krause, T. O. Wehling, S. Lounis, A. Lichtenstein, D. Pfannkuche, J. Wiebe, and R. Wiesendanger, Science \textbf{339}, 55 (2013).

\bibitem{Parkin:2008gs} S. S. P. Parkin, M. Hayashi, and L. Thomas, Science \textbf{320}, 190 (2008).

\bibitem{Thomas:2010ta} L. Thomas, R. Moriya, C. Rettner, and S. S. Parkin, Science \textbf{330}, 1810 (2010).

\bibitem{Bauer:2012fq} G. E. W. Bauer, E. Saitoh, and B. J. van Wees, Nature Materials \textbf{11}, 391 (2012).

\bibitem{Ralph:2008kj} D. C. Ralph and M. D. Stiles, J. Mag. and Magnetic Materials \textbf{320}, 1190 (2008).

\bibitem{Brataas:2012fb} A. Brataas, A. D. Kent, and H. Ohno, Nature Materials 11, 372 (2012).

\bibitem{Zhang:2004hs} S. Zhang and Z. Li, Phys. Rev. Lett. \textbf{93}, 127204 (2004).

\bibitem{Tatara:2004ub} G. Tatara and H. Kohno, Phys. Rev. Lett. \textbf{92}, 086601 (2004).

\bibitem{SSE-metal} K. Uchida, S. Takahashi, K. Harii, J. Ieda, W. Koshibae, K. Ando, S. Maekawa, E. Saitoh, Nature \textbf{455}, 778 (2008).
	
\bibitem{SSE-inst}	K. Uchida, J. Xiao, H. Adachi, J. Ohe, S. Takahashi, J. Ieda, T. Ota, Y. Kajiwara, H. Umezawa, H. Ka-wai, G. E.W. Bauer, S. Maekawa and E. Saitoh, Nature Mat.  \textbf{9}, 894  (2010).

\bibitem{SSE-semi}	 C. M. Jaworski, J. Yang, S. Mack, D. D. Awschalom, J. P. Heremans and R. C. Myers, Nature Mat.  \textbf{9}, 898 (2010).

\bibitem{SSE-nom} C. M. Jaworski, R. C. Myers, E. Johnston-Halperin, and J. P. Heremans, Nature \textbf{487}, 210 (2012).

\bibitem{Kajiwara:2010uc}	Y. Kajiwara, K. Harii, S. Takahashi, J. Ohe, K. Uchida, M. Mizuguchi, H. Umezawa, H. Kawai, K. Ando, and K. Takanashi, Nature 464, 262 (2010).

\bibitem{Shikoh:2013jp} E. Shikoh, K. Ando, K. Kubo, E. Saitoh, T. Shinjo, and M. Shiraishi, Phys. Rev. Lett. \textbf{110}, 127201 (2013).

\bibitem{Book-SpinCurrent} S. Maekawa,  Sergio O. Valenzuela, E. Saitoh, and T. Kimura, \emph{Spin Current} (Oxford University Press, 2012).

\bibitem{DWs-JB06} V.K. Dugaev, J. Berakdar, and J. Barna\'{s}, Phys. Rev. Lett. \textbf{96}, 047208 (2006).


\bibitem{SSE-ansatz} C.L. Jia and J. Berakdar, Phys. Rev. B \textbf{83}, 180401(R) (2011).

\bibitem{Barnes:2007ko} S. Barnes and S. Maekawa, Phys. Rev. Lett. \textbf{98}, 246601 (2007).

\bibitem{Yang:2009ko} S. Yang, G. Beach, C. Knutson, D. Xiao, Q. Niu, M. Tsoi, and J. Erskine, Phys. Rev. Lett. \textbf{102}, 067201 (2009).

\bibitem{Schulz:2012bi} T. Schulz, R. Ritz, A. Bauer, M. Halder, M. Wagner, C. Franz, C. Pfleiderer, K. Everschor, M. Garst, and A. Rosch, Nature Physics \textbf{8}, 301 (2012).


\bibitem{Dugaev:2006ed} V. Dugaev, V. Vieira, P. Sacramento, J. Barnas, M. AraÂjo, and J. Berakdar, Phys. Rev. B \textbf{74}, 054403 (2006).


\bibitem{Uchida:2012ew} K. Uchida, T. Ota, H. Adachi, J. Xiao, T. Nonaka, Y. Kajiwara, G. Bauer, S. Maekawa, and E. Saitoh, J. Appl. Phys. \textbf{111}, 103903 (2012).


\bibitem{Saitoh:2004fa} E. Saitoh, H. Miyajima, T. Yamaoka, and G. Tatara, Nature \textbf{432}, 203 (2004).

\bibitem{Book-Yosida} K. Yosida, \emph{Theory of Magnetism},  (Springer-Verlag, Berlin Heidelberg, 1996).

\bibitem{JOHN} C.L. Jia, S. Onoda, N. Nagaosa, and J.H. Han, Phys. Rev. B \textbf{74}, 224444 (2006); Phys. Rev. B \textbf{76}, 144424 (2007).

\bibitem{p-NE-1} Yong Pu, E. Johnston-Halperin, D. D. Awschalom, and Jing Shi, Phys. Rev. Lett. \textbf{97}, 036601 (2006).

\bibitem{p-NE-2} A. D. Avery, M. R. Pufall, and B. L. Zink, Phys. Rev. Lett. \textbf{109}, 196602 (2012).

\bibitem{Bauer:2010iv} G. E. W. Bauer, S. Bretzel, A. Brataas, and Y. Tserkovnyak, Phys. Rev. B \textbf{81}, 024427 (2010).

\bibitem{Bazaliy:1998vo} Y. B. Bazaliy, B. A. Jones, and S.-C. Zhang, Phys. Rev. B \textbf{57}, R3213 (1998).

\bibitem{Miron:2011fn} I. M. Miron, T. Moore, H. Szambolics, L. D. Buda-Prejbeanu, S. Auffret, B. Rodmacq, S. Pizzini, J. Vogel, M. Bonfim, and A. Schuhl, Nature Materials \textbf{10}, 419 (2011).

\bibitem{Curiale:2012iq} J. Curiale, A. Lematre, C. Ulysse, G. Faini, and V. Jeudy, Phys. Rev. Lett. \textbf{108}, 076604 (2012).

\bibitem{Alexander:2009rl} A. Sukhov, J. Berakdar, Phys. Rev. Lett. \textbf{102}, 057204 (2009).

\bibitem{footnt} The calculations were performed self-consistently while taking into account simultaneously the development of the out-of-plane components of the second DW, \emph{i.e.}, the polar angle $\alpha$ is relaxed to vary with $t$ during the LLG dynamics.

\bibitem{Li:2004kh} Z. Li and S. Zhang, Phys. Rev. Lett. \textbf{92}, 207203 (2004).


\bibitem{Jung:2008kq} S.-W. Jung, W. Kim, T.-D. Lee, K.-J. Lee, and H.-W. Lee, Appl. Phys. Lett. 92, 202508 (2008).

\bibitem{Mostovoy:2011dz} M. Mostovoy, K. Nomura, and N. Nagaosa, Phys. Rev. Lett. \textbf{106}, 047204 (2011).


\bibitem{metals}
 For metals, DW spatial variation  is larger than $2\pi/k_{F}$.  Scattering in  treated perturbatively using the Green's function constructed from Eq.(\ref{eq:psi}) (cf. Refs.~[\citenum{DWs-Nich}] and [\citenum{JB-jpa03}]).   Effects of non-adiabatic transfer torque, the so called $\beta$ term  emerge and  may induce a steady state DW motion \cite{Zhang:2004hs,Curiale:2012iq}.
 %

\bibitem{DWs-Nich} N. Sedlmayr, V. K. Dugaev, and J. Berakdar, Phys. Rev. B \textbf{79}, 174422 (2009); N. Sedlmayr, V. K. Dugaev, M. Inglot, and J. Berakdar, Phys. Status Solidi RRL \textbf{5}, 450 (2011).

%
\bibitem{JB-jpa03} V. K. Dugaev, J. Barnas and J. Berakdar, J. Phys. A  \textbf{36}, 9263 (2003).

\bibitem{Gourdon:2007ko} C. Gourdon, A. Dourlat, V. Jeudy, K. Khazen, H. von Bardeleben, L. Thevenard, and A. Lematre, Phys. Rev. B \textbf{76}, 241301(R) (2007).

\bibitem{Matsukura:1998tx} F. Matsukura, H. Ohno, A. Shen, and Y. Sugawara, Phys. Rev. B \textbf{57}, R2037 (1998).


\bibitem{Goennenwein:2003kl} S. T. B. Goennenwein, T. Graf, T. Wassner, M. S. Brandt, M. Stutzmann, J. B. Philipp, R. Gross, M. Krieger, K. Zürn, P. Ziemann, A. Koeder, S. Frank, W. Schoch, and A. Waag, Appl. Phys. Lett. \textbf{82}, 730 (2003).

\bibitem{Haghgoo:2010ik} S. Haghgoo, M. Cubukcu, H. J. von Bardeleben, L. Thevenard, A. Lemaitre, and C. Gourdon, Phys. Rev. B \textbf{82}, 041301(R) (2010).

\bibitem{Dourlat:2008gw} A. Dourlat, V. Jeudy, A. Lematre, and C. Gourdon, Phys. Rev. B \textbf{78}, 161303(R) (2008).

\bibitem{Adam:2009iu} J.-P. Adam, N. Vernier, J. Ferré, A. Thiaville, V. Jeudy, A. Lematre, L. Thevenard, and G. Faini, Phys. Rev. B \textbf{80}, 193204 (2009).


\bibitem{Khazen:2008fw} K. Khazen, H. J. von Bardeleben, M. Cubukcu, J. L. Cantin, V. Novak, K. Olejnik, M. Cukr, L. Thevenard, and A. Lematre, Phys. Rev. B \textbf{78}, 195210 (2008).

\bibitem{Cubukcu:2010cy} M. Cubukcu, H. J. von Bardeleben, K. Khazen, J. L. Cantin, O. Mauguin, L. Largeau, and A. Lematre, Phys. Rev. B \textbf{81}, 041202 (2010).

\bibitem{Mn12} One FM could  be a heated, unbiased STM tip. Other FM serves as a substrate for Mn$_{12}$ molecules \cite{Heersche:2006cs, Jo:2006ck}.
 The total spin and diameter of Mn$_{12}$ are $M_0 =10$ and $w=3$ nm; the exchange coupling is $gM_0 = 1$ meV \cite{Wang:2010gn}, and $\alpha_{G} \approx 0.17$ and $\hbar/g \approx 6.6 \times 10^{-12} s$. These parameters are in the range of our simulations.

%\bibitem{Yuan:2010hb} Z. Yuan, S. Wang, and K. Xia, Solid State Commun. \textbf{150}, 548 (2010).

%\bibitem{Hals:2010eu} K. M. D. Hals, A. Brataas, and G. E. W. Bauer, Solid State Commun. \textbf{150}, 461 (2010).

\bibitem{Heersche:2006cs} H. B. Heersche, Z. de Groot, J. A. Folk, H. Van der Zant, C. Romeike, M. R. Wegewijs, L. Zobbi, D. Barreca, E. Tondello, and A. Cornia, Phys. Rev. Lett. \textbf{96}, 206801 (2006).

\bibitem{Jo:2006ck} M.-H. Jo, J. E. Grose, K. Baheti, M. M. Deshmukh, J. J. Sokol, E. M. Rumberger, D. N. Hendrickson, J. R. Long, H. Park, and D. C. Ralph, Nano Lett. \textbf{6}, 2014 (2006).


\bibitem{Wang:2010gn} R.-Q. Wang, L. Sheng, R. Shen, B. Wang, and D. Y. Xing, Phys. Rev. Lett. \textbf{105}, 057202 (2010).

%
\end{thebibliography}
\end{document}